\documentclass[screen, nonacm]{acmart}

\AtBeginDocument{%
  }

\renewcommand\footnotetextcopyrightpermission[1]{} 
\acmConference[Conference acronym 'XX]{Make sure to enter the correct
  conference title from your rights confirmation emai}{June 03--05,
  2018}{Woodstock, NY}
\acmISBN{978-1-4503-XXXX-X/18/06}

\usepackage{listings}
\lstdefinestyle{base}{
  emptylines=1,
  moredelim=**[is][\color{red}]{@}{@},
}
\usepackage{draftwatermark}

\SetWatermarkText{PREPRINT}
\SetWatermarkColor[gray]{0.5}
\SetWatermarkFontSize{1cm}
\SetWatermarkAngle{90}
\SetWatermarkHorCenter{20cm}
\usepackage{booktabs}
\usepackage{multirow}
\usepackage[normalem]{ulem}
\useunder{\uline}{\ul}{}

\begin{document}

\title{PAIGE: Examining Learning Outcomes and Experiences with Personalized AI-Generated Educational Podcasts}

\author{Tiffany D. Do}
\email{tiffany.do@ucf.edu}
\affiliation{%
  \institution{Drexel University and Google}
  \city{Philadelphia}
  \country{United States}
}

\author{Usama Bin Shafqat}
\affiliation{%
  \institution{Google}
  \city{Mountain View}
  \country{United States}
}

\author{Elsie Ling}
\affiliation{%
  \institution{Google}
  \city{Mountain View}
  \country{United States}
}

\author{Nikhil Sarda}
\email{nikhilsarda@google.com}
\affiliation{%
  \institution{Google}
  \city{Mountain View}
  \country{United States}
}
\renewcommand{\shortauthors}{Do et al.}

\begin{abstract}
  Generative AI is revolutionizing content creation and has the potential to enable real-time, personalized educational experiences. We investigated the effectiveness of converting textbook chapters into AI-generated podcasts and explored the impact of personalizing these podcasts for individual learner profiles. We conducted a 3x3 user study with 180 college students in the United States, comparing traditional textbook reading with both generalized and personalized AI-generated podcasts across three textbook subjects. The personalized podcasts were tailored to students’ majors, interests, and learning styles. Our findings show that students found the AI-generated podcast format to be more enjoyable than textbooks and that personalized podcasts led to significantly improved learning outcomes, although this was subject-specific. These results highlight that AI-generated podcasts can offer an engaging and effective modality transformation of textbook material, with personalization enhancing content relevance. We conclude with design recommendations for leveraging AI in education, informed by student feedback.
\end{abstract}

\ccsdesc[500]{Human-centered computing~User studies}
\ccsdesc[500]{Applied computing~Education}

\keywords{artificial intelligence in education, personalized learning, large language models, content transformation}


\maketitle

\section{Introduction}
In the era of digital learning, podcasts are rapidly becoming essential tools for higher education across various disciplines, including language \cite{chan2011students}, medicine \cite{kellyLearning2022}, and marketing \cite{mccarthy2021}. For instance, a review by Shahrizal et al. \cite{shahrizal2022systematic} found that university students increasingly prefer podcasts as a study method. Similarly, Kelly et al. \cite{kellyLearning2022} report that podcasts have gained widespread acceptance in medical education, with most medical residents adopting them as a key learning resource.

Conversely, traditional textbooks are becoming increasingly challenging for students to engage with as learning materials. Baron and Mangen \cite{baron2021doing} observe that while textbooks remain essential in humanities and arts education, their role as a primary learning modality is declining, partly due to the growing time higher education students spend on electronic media and other activities. Similarly, Gorzycki et al. \cite{gorzyckiReading2020} found that although college students recognize the importance of textbooks, they often fail to complete their reading assignments, citing perceptions of "irrelevance" and "boredom." Given these challenges, there is a clear need for more engaging educational materials. Previous studies have shown that students not only enjoy but also learn more effectively from podcasts based on textbook content compared to traditional reading methods \cite{back2017, evans2008}. However, the process of converting textbook chapters into podcasts is often tedious for educators, limiting its broader application.

As generative artificial intelligence (AI) continues to advance, there is increasing potential to address this gap by transforming traditional textbooks in real time into more engaging formats, such as podcasts, that better align with students' preferences. Moreover, the capabilities of large language models (LLMs) offer the opportunity to tailor these lessons to each student's unique learning needs, preferences, and prior knowledge. This level of personalization can enhance the relevance and entertainment of educational content, addressing the negative perceptions associated with traditional textbooks. By ensuring that material resonates more closely with individual students' interests and needs, LLMs have the potential to make learning more meaningful and engaging in real time.

In this paper, we explore the use of generative AI to convert textbook chapters into educational podcasts using advanced language and voice models. Building on the demonstrated effectiveness of human-generated educational podcasts compared to traditional textbooks \cite{back2017, evans2008}, we investigate whether AI-generated podcasts, particularly those personalized to student interests and majors, can similarly enhance the learning experience. We address the following research questions:

\begin{itemize}
    \item \textbf{RQ1:} How do students’ preferences for AI-generated podcasts compare to those for traditional textbooks?
    \item \textbf{RQ2:} What impact do AI-generated podcasts have on students' learning outcomes?
    \item \textbf{RQ3:} How do variations in personalization and subject matter influence students' engagement and learning outcomes?
\end{itemize}

To answer these questions, we conducted an exploratory user study involving current college students (n=180) in the United States, who either read a textbook chapter, listened to a personalized podcast, or listened to a generalized podcast. We examined learning outcomes, enjoyment, and engagement, and gathered qualitative feedback. This paper makes the primary contributions: 
\begin{enumerate}
    \item We introduce a method for converting textbook chapters into AI-generated podcasts, offering a novel way to engage with academic content.
    \item Through a user study with 180 college students, we provide evidence that AI-generated podcasts are more enjoyable than textbooks, with personalized versions further improving effectiveness.
    \item Our study reveals that personalizing podcasts to students’ majors and interests can significantly enhance learning outcomes, though effects vary by subject.
\end{enumerate}
\section{Related Work}
\subsection{Personalized Learning Experiences}
Personalized learning has long intrigued the educational community and is widely recognized as a significant 21st-century challenge, whether approached through human-driven or automated methods \cite{bernackiSystematic2021, brass2020personalized}. A recent review by Li and Wong \cite{liPersonalisation2023} reported that most technology-based personalization efforts focus on dynamically adjusting content difficulty based on learners' abilities. Moreover, much of the research on AI in this area has targeted personalizing assessment outcomes and instructional strategies, such as adjusting task difficulty or providing tailored feedback (e.g., \cite{Park2024, Antonova2019}) or offering personalized recommendations (e.g., \cite{Perez-Ortiz2021}). In a recent review, Chaudry and Kazim \cite{chaudhry2022artificial} state that current work in contextualized learning helps with "identifying the learning gaps in each learner, offer content recommendations based on that and provide step by step solutions to complex problems".

In contrast, our focus is on personalized \textit{context}, which involves adapting educational materials to align with a student's interests, leveraging situational interest and prior knowledge \cite{bernacki2018role}. A comprehensive 2024 systematic review by Lin et al. \cite{linPersonalized2024} found that interest-based personalization had a moderate effect on cognitive load and knowledge transfer across several studies. For instance, an early study demonstrated that AI systems could select reading passages based on student interests, positively impacting learning outcomes \cite{heilmanPersonalization2010}. However, some studies suggest that interest-based personalization may not always aid learning and can sometimes reduce enjoyment \cite{iterbeke2022role, VANDEWEIJERBERGSMA2021}. It is important to note that these studies focused on selection-based personalization, where either a student or system chooses among predefined contexts, rather than generating personalized content in real time.

Lin et al. \cite{linPersonalized2024} highlight that much of the existing research has relied on paper-based learning methods and has yet to explore the potential of generative AI for creating personalized educational content. As generative AI becomes more prevalent, its integration into educational tools becomes increasingly relevant for Human-Computer Interaction (HCI). Generative AI offers the potential to create and personalize context-aware educational content, particularly in audio formats, which has not yet been explored in current research.

Walkington and Bernacki \cite{WalkingtonBernacki2020} identify several persistent challenges in personalized learning, including technology integration, learner agency, and the lack of consistent theoretical grounding. To address these challenges, we developed and evaluated one of the first systems utilizing generative AI for personalized learning. Our system enables learners to customize the level of personalization, giving them control over the amount of information included. Additionally, our approach is grounded in textbook content, ensuring factual accuracy.

\subsection{AI-Generated Educational Content}
The use of generative AI for educational content has gained prominence in recent years with the advancement and widespread adoption of this technology. Multiple studies and reviews have found that university students hold positive opinions about using generative AI in education and are increasingly leveraging it for studying and learning purposes \cite{abdaljaleel2024multinational, ngo2023perception, ALBAYATI2024100203}. For instance, a recent study by Denny et al. \cite{denny2023trust} reported that university students found AI-generated educational content to be as helpful as human-generated content. Similarly, Han et al. \cite{Han2024} found that teachers of younger students viewed generative AI as a valuable tool for creating adaptable teaching materials, reflecting a growing acceptance of AI in education. As the educational landscape rapidly evolves, it is important to understand how to best harness this technology to create effective learning materials for students.

Several scholars have explored using generative AI to simulate interactive instructors, a rapidly evolving field in recent years. For instance, Chen et al. \cite{chen2023empowering} investigated LLM-powered agents for explaining concepts and answering questions, finding that LLMs can effectively serve as intelligent tutors. Kahl et al. \cite{kahl2024evaluating} found that an LLM tutor using Retrieval-Augmented Generation (RAG) improved model responses and factual answers, providing support for the idea of transforming content, such as textbook materials, in real time. However, most of these agents have focused on roles such as answering student questions (e.g., \cite{Tu2023, sonkar-etal-2023-class, kahl2024evaluating, chen2023empowering, ALSAFARI2024100101}) or providing support in a learning environment (e.g., \cite{Wang2024}). Our system, in contrast, aims to enhance traditional learning materials by transforming them into a more engaging and personable format.

Although creating podcasts from text content is a relatively new field, prior work has explored using podcasts or conversational agents for news. For example, Laban et al. \cite{laban2022} found that AI-generated news podcasts were effective in helping university students stay informed about current events, with most students responding positively. Nordberg and Guribye \cite{nordbergguribye2023} examined conversational agents for news and noted that these agents can assume roles ranging from assistant to companion. We build upon this research by incorporating user information, allowing the agent to adopt a role between educator and companion through content tailoring. While these previous studies have focused on news education, our work concentrates on transforming textbook content.

\section{Personalized AI-Generated Educational (PAIGE) Podcast System}
This section introduces our system for generating podcasts from textbook chapters. We begin by outlining the context data used, followed by a description of the text transcript generation process. Finally, we conclude with a discussion of the voice models employed for text-to-speech (TTS) synthesis. Samples of podcast generations can be found in our supplementary materials.

\subsection{Podcast Context Data}
To generate the podcasts, we utilized open source textbooks from OpenStax\footnote{We received explicit permission from OpenStax to use the textbooks for this study.}. Chapters from these textbooks served as grounding data, with the prompts instructing the model to create a podcast that taught the chapter content. For Personalized podcasts, we included user profile information as contextual data, encompassing their major, age group, interests, and preferred learning style. This information was collected through a survey administered prior to the study.

\subsection{Transcript Generation}

\begin{figure}
  \includegraphics[width=\textwidth]{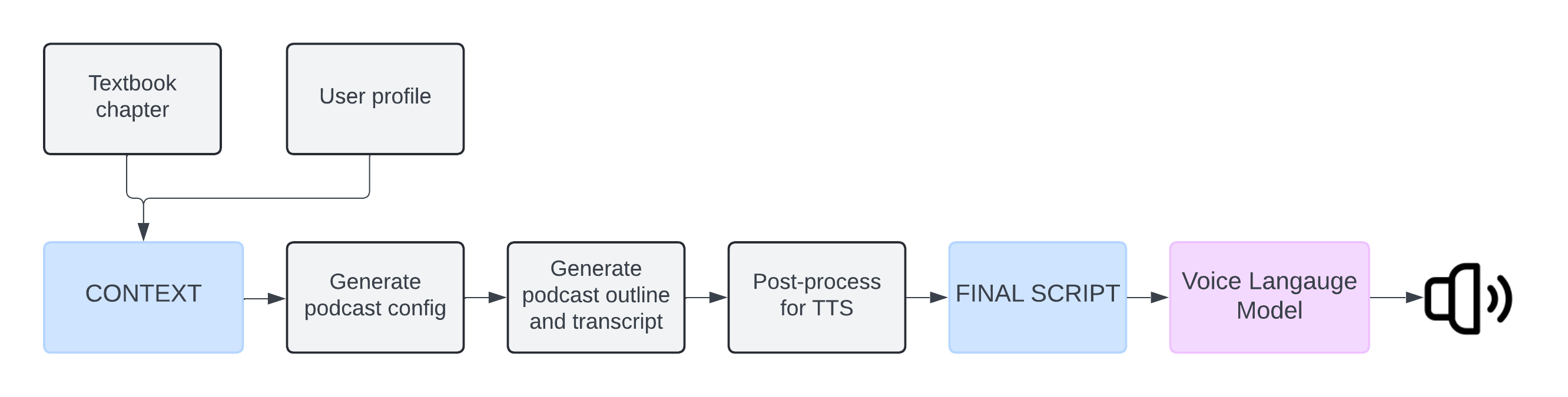}
  \caption{A flowchart of the podcast generation pipeline.}
  \Description{A flowchart showing the generation pipeline. First the context, which is the chapter and profile, is provided. Then, the config, outline, and transcript is generated.}
  \label{fig:pipeline}
\end{figure}
We employed a multi-part generation pipeline using Gemini 1.5 Pro to generate podcast transcripts (see Figure \ref{fig:pipeline}). Our steps for generating a transcript are as follows: 

\begin{enumerate}
    \item Configuration: A configuration file is generated for the textbook content, which includes details such as the host and expert's personas and the podcast's content. For personalized podcasts, the configuration is also generated from the user profile, tailoring the transcript to their individual learning needs.
    \item Outline: An outline of the podcast is generated using the configuration file and podcast data (textbook content and user profile, if personalized). This outline step was based on the "Skeleton of Thought" concept, which "first guides LLMs to generate the skeleton of the answer" \cite{ning2024skeletonofthought} to improve processing time through parallel generation and enhance quality. Our preliminary pilot studies confirmed the benefit of outlines in enhancing transcript quality.
    \item Transcript: A full transcript is generated using the outline. Afterwards, this transcript is fact-checked against the textbook source material. If factual errors exist, the script is modified until no factual errors are detected. 
    \item Post-processing for TTS: The transcript is post-processed to ensure that it can be passed through a voice language model. This step includes removing any artifacts, such as parenthesis or miscellaneous puncutation.
\end{enumerate}

\begin{figure}
\noindent\begin{minipage}{.45\textwidth}
\begin{lstlisting}[frame=single,style=base,caption=A generalized transcript for a college audience]{Generalized}
Host: So, we've got primary and secondary reinforcers. What's the difference?
Guest: Well, primary reinforcers are like, the inherently good stuff. Think of food, water, or a good night's sleep. You don't need to learn to enjoy them, they just feel good.
Host: Got it, like that satisfaction after a good meal.
Guest: Exactly! Now, secondary reinforcers are things that become rewarding because they're linked to those primary things.
Host: Like money?
Guest: Yep! Money itself isn't inherently good, but it lets you buy things that are good, like food, experiences, or a comfortable place to sleep. 
\end{lstlisting}
\end{minipage}\hfill
\begin{minipage}{.45\textwidth}
\begin{lstlisting}[frame=single,style=base,caption=A personalized transcript for a student who has an interest in K-pop]{Personalized} 
Guest: Well, we've got our primary reinforcers and our secondary reinforcers. @Let's say you finally managed to snag those concert tickets to see [KPOP BAND].@
Host: Dream come true.
Guest: Right. @So, the actual experience itself the music, the energy, the excitement of seeing them live that's a primary reinforcer.@ It's inherently rewarding.
Host: Okay, so then secondary reinforcers those are the things that become rewarding because they're connected to those primary rewards, right.
Guest: Exactly. Think of money, for example. It doesn't have inherent value on its own, but @it allows you to buy those concert tickets, the merch, the albums@ all those things that get you closer to the primary reward.
\end{lstlisting}
\end{minipage}
\caption{Excerpts of generalized and personalized transcripts that explain the concept of primary and secondary psychological reinforcers.}
    \label{fig:configs}
\end{figure}

\begin{figure}[h!] \centering
\includegraphics[width=\textwidth]{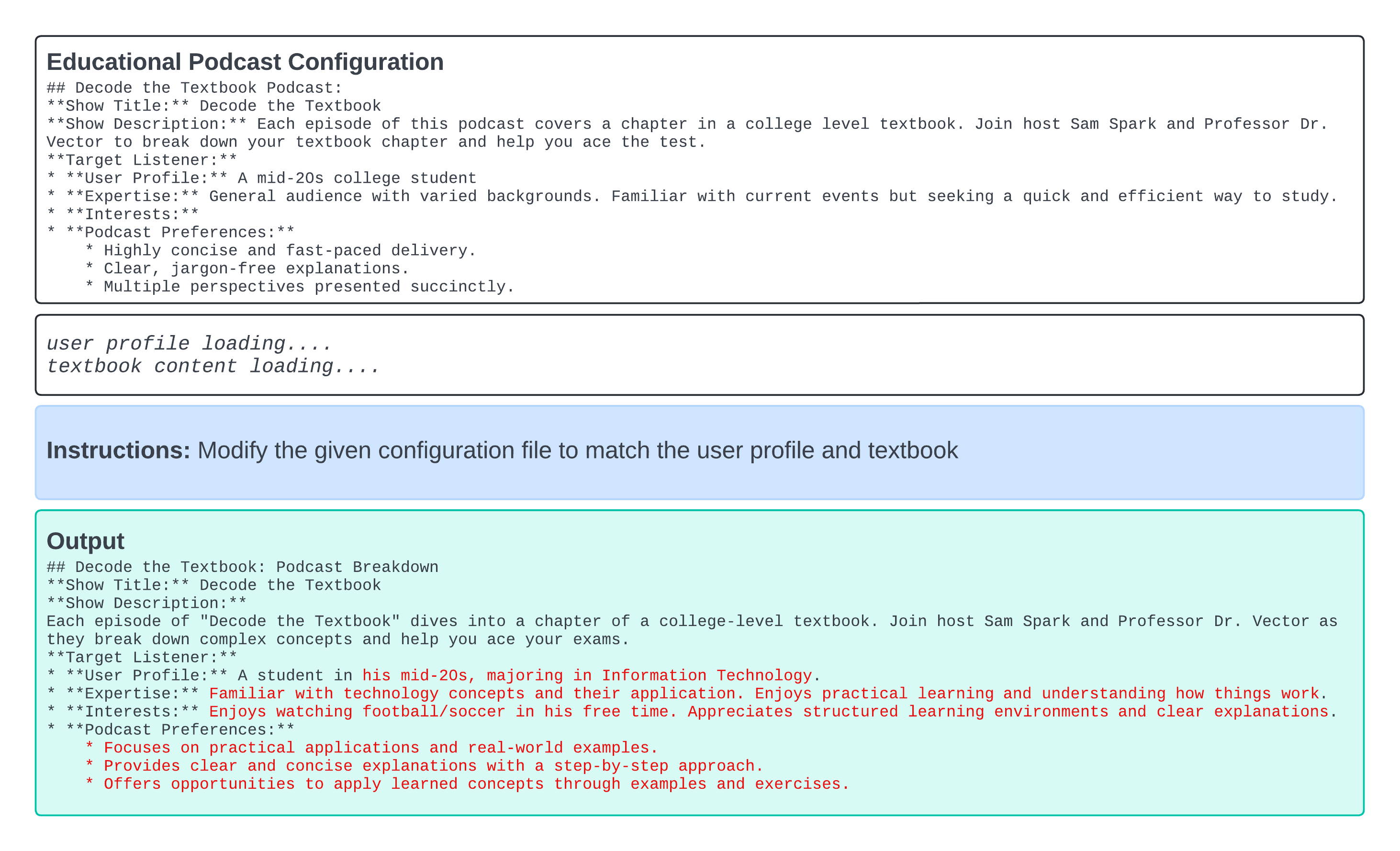}
\Description{Example of configuration generation for a personalized podcast. A given configuration format is modified to fit the user profile, which includes elements such as the user's age, major, and podcast preferences.}
\caption{Example of configuration generation for a personalized podcast. A given configuration format is modified to fit the user profile. Text in red shows modifications to the original podcast configuration.}
\label{fig:config}
\end{figure}

\subsection{Speech Generation}
To achieve high-quality Text-to-Speech (TTS), we generated audio using a TTS voice language model based on AudioLM \cite{audiolm,kharitonov2023speak}. This model was chosen for its ability to efficiently generate speech that closely matches the naturalness and acoustic quality of ground-truth recordings \cite{kharitonov2023speak}. To maintain audio consistency across all generated podcasts, we used the same two speakers, a male and a female voice, both licensed from young adult voice actors. Voice actors had American English accents, which is optimal for TTS instructors for students in the United States \cite{Do2022}.

\section{Educational Podcast Comparison Study}
We conducted a study to compare learning outcomes and learning experience between personalized AI-generated podcasts (\textit{Personalized}) and AI-generated generalized podcasts (\textit{Generalized}). In addition, we included textbook reading of the original textbook content (\textit{Textbook}) as a baseline comparison

This study was conducted using three OpenStax college textbooks: "Introduction to Philosophy" (Chapter 1) \cite{smith_2022}, "Psychology 2e" (Chapter 6) \cite{psych_book}, and "American Government 3e" (Chapter 1) \cite{krutzAmerican2021}. We selected these chapters for their comparable length and because they did not require prior knowledge. Written permission from OpenStax was obtained to use the textbooks for this study.

\subsection{Research Hypotheses}
We present the following research hypotheses in the context of AI-generated podcasts:
\\
\textbf{H1:} Based on personalization theory \cite{bernacki2018role}, we hypothesize that participants will rate the \textit{Personalized} condition as providing the most positive learning experience. Additionally, drawing on prior work demonstrating the engaging and enjoyable nature of podcasts compared to textbooks \cite{back2017, evans2008}, we expect the \textit{Generalized} condition to receive higher ratings than the \textit{Textbook} condition.
\\ \\
\textbf{H2:} We expect that the \textit{Personalized} podcasts will lead to higher learning outcomes than the \textit{Generalized} podcasts, due to prior research indicating that interest-based personalization positively affects learning outcomes \cite{linPersonalized2024}.
\\ \\
\textbf{H3:} We do not expect any differences based on textbook subject, due to a lack of prior work in this area.

\subsection{Dependent Variables}
Full questionnaires administered can be found in our supplemental materials \footnote{We do not provide the questions and answers for the knowledge retention questionnaires due to OpenStax policy. Verified educators from academic institutions may freely access test banks through OpenStax.}. 

\subsubsection{Learning Outcomes}
To assess knowledge retention, participants completed multiple-choice tests consisting of ten questions randomly selected from the textbook's test bank. These questions, drawn from each chapter subsection in roughly equal proportions, were graded using the test bank's answer key. Participants' scores ranged from 0 to 10. The podcast generator models were unaware of the test questions beforehand to ensure an unbiased assessment.

\subsubsection{Learning Experience}
To assess user experience, we employed the User Experience Questionnaire (UEQ) \cite{laugwitz2008construction}, a widely used tool for evaluating educational consumer product satisfaction. We focused on two UEQ subscales: Attractiveness (measuring overall user appeal) and Stimulation (assessing user engagement and motivation). These subscales were chosen to gauge the potential of generative AI podcasts for educational purposes. The UEX uses a 7-point anchored Likert scale, with anchors such as "Annoying-Enjoyable" for \textit{Attractiveness} and "Demotivating-Motivating" for \textit{Stimulation}.

\subsection{Procedure}
The study consisted of a single 35-minute remote session. Participants were recruited through the Prolific online marketplace. Prior to the session, participants completed a survey that collected information on their age group, major, interests, and preferred teaching styles. The survey included open-ended questions, allowing participants to describe their profiles with as much detail as they preferred. Informed consent for data collection was obtained in accordance with our institution's policies, and participants were informed that their data would be used to generate educational content. 

At the start of the study, participants were directed to a custom web-based survey to complete their assigned condition. They were instructed to "Study the given material to prepare for a quiz." For those in the podcast conditions, participants were required by the system to listen to the entire podcast before proceeding to the next section. Due to the nondeterministic nature of generative AI, the length of the podcasts varied slightly with each generation. To maintain consistency, we limited the podcasts to a duration of 10-12 minutes, with an average length of 11 minutes. Participants in the textbook condition viewed the chapter in a web-based PDF reader and were similarly required to spend 11 minutes on the task, ensuring uniformity across all conditions.

Participants then completed the UEQ questionnaire (5 minutes) and the knowledge test (10 minutes). Finally, they provided freeform responses, describing what they liked and disliked about the experience (5 minutes). Participants were compensated with \$7 USD through the Prolific platform.

\subsection{Participants}
\begin{figure}[h!] \centering
  \includegraphics[width=5in]{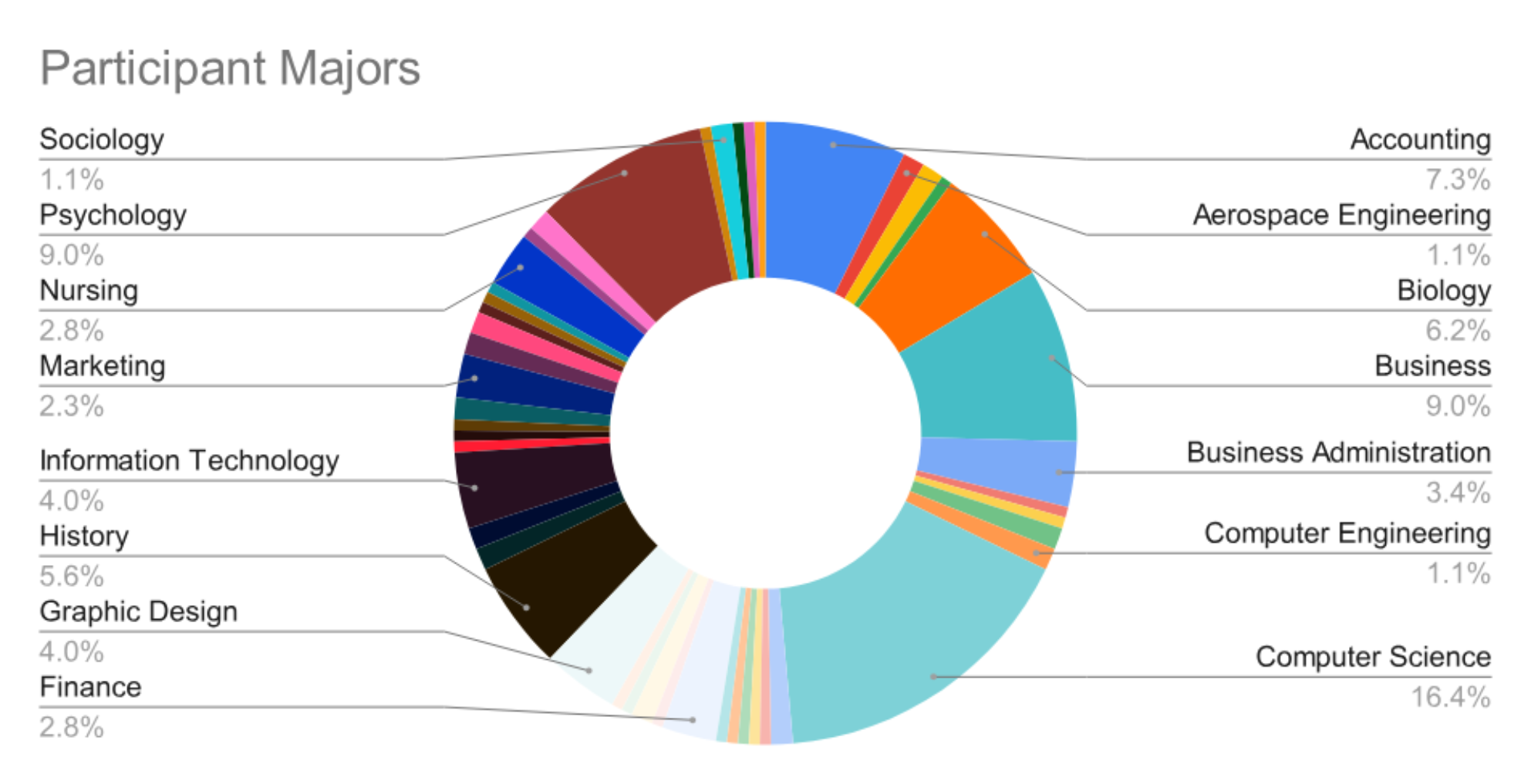}
  \caption{Participant major distribution. Detailed demographic breakdown is displayed in Table \ref{table:participants}.}
  \Description{A figure showing the distribution of participant majors. Majors are distributed across a wide variety, including Computer Science, Psychology, and Business.}
  \label{fig:majors}
\end{figure}

We recruited 180 current college students in the United States, with 60 participants randomly assigned to each of the textbook subjects. Participants were assigned to one of the three content style conditions: \textit{Generalized}, \textit{Personalized}, and \textit{Textbook}. All participants were pre-screened to ensure they had not taken the relevant course and reported minimal prior knowledge in the subject area. Table \ref{table:participants} provides a detailed breakdown of participant demographics (age, gender, major of study) across conditions. Figure \ref{fig:majors} shows an infographic of the major breakdown of all participants.

\subsection{Data Analysis Approach}
We measured \textit{Attractiveness} and \textit{Stimulation} by averaging the scores across all items in each respective scale. Chapter test scores were computed using the test bank. To analyze the data, we conducted a 3x3 ANOVA to examine the interaction and main effects of content style (\textit{Generalized}, \textit{Personalized}, and \textit{Textbook}) and textbook subject (Philosophy, Psychology, and Government). Post hoc analysis was conducted using Tukey's HSD (Honestly Significant Difference) tests to compare pairwise differences between group means.

\section{Results}
The means and standard deviations, organized by subject, can be found in Table \ref{table:results}. The results of all ANOVA tests for main and interaction effects can be found in Table \ref{table:anova}.

\begin{table}[]
\caption{Means (M) and standard deviations (SD) of all dependent variables, disaggregated by subject and content style.}
\begin{tabular}{|c|c|ccc|}
\hline
\multicolumn{1}{|l|}{} & \multicolumn{1}{l|}{\textbf{Content Type}} & \textbf{\begin{tabular}[c]{@{}c@{}}Test Score\\ M (SD)\end{tabular}} & \textbf{\begin{tabular}[c]{@{}c@{}}Attractiv.\\ M (SD)\end{tabular}} & \textbf{\begin{tabular}[c]{@{}c@{}}Stimulation\\ M (SD)\end{tabular}} \\ \hline
\multirow{3}{*}{{\textbf{Philosophy}}} & \textbf{Generalized} & 5.85 (1.35) & 5.26 (0.67) & 4.86 (1.05) \\
 & \textbf{Personalized} & 7.30 (1.45) & 5.15 (0.84) & 4.95 (1.05) \\
 & \textbf{Textbook} & 5.95 (1.57) & 4.58 (1.23) & 4.42 (1.31) \\ \hline
\multirow{3}{*}{{\textbf{Psychology}}} & \textbf{Generalized} & 4.75 (1.62) & 5.29 (1.17) & 4.96 (1.34) \\
 & \textbf{Personalized} & 5.70 (1.38) & 5.63 (1.06) & 5.32 (1.24) \\
 & \textbf{Textbook} & 4.80 (1.57) & 4.30 (1.05) & 4.45 (1.04) \\ \hline
\multirow{3}{*}{{\textbf{Government}}} & \textbf{Generalized} & 6.10 (1.45) & 5.22 (0.75) & 4.63 (0.81) \\
 & \textbf{Personalized} & 6.70 (1.38) & 4.98 (1.19) & 4.64 (1.25) \\
 & \textbf{Textbook} & 7.15 (1.27) & 4.67 (0.68) & 4.68 (0.89) \\ \hline
\end{tabular}
\label{table:results}
\end{table}

\subsection{Learning Outcomes}
\begin{figure} \centering
  \includegraphics[width=\textwidth]{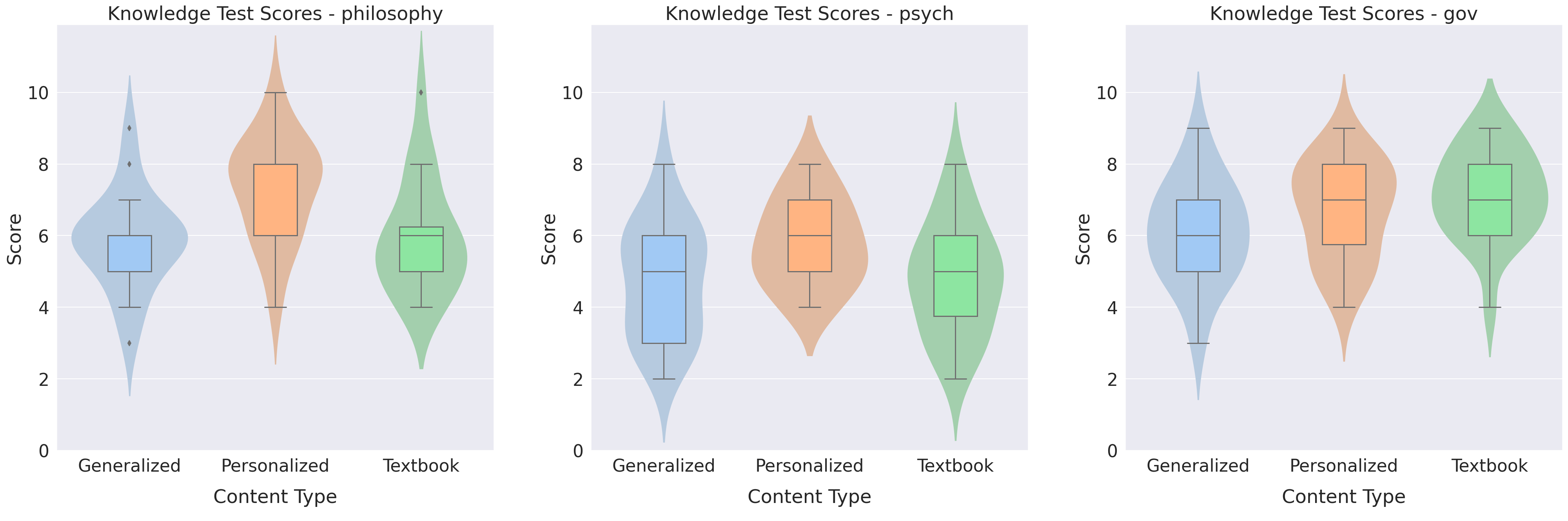}
  \caption{Violin and boxplots of all test scores by subject and content style condition.}
  \Description{Violin and boxplots of all test scores by subject. For philosophy and psychology, the score for personalized podcasts is significantly higher than the other conditions, although the difference is smaller for psychology. There is no significant difference of content type for government.}
  \label{fig:test_scores}
\end{figure}

Contrary to H3, we found a significant interaction effect between \textit{content style} and \textit{textbook subject}, $F(4, 171)=2.27, p <0.01$, which indicated that we should analyze scores of each subject separately. Plots for all subjects are displayed in Figure \ref{fig:test_scores}. 
\begin{itemize}
    \item \textbf{Philosophy}: We found a significant main effect of \textit{content style}, $F(2,57)= 6.14, p<0.01$ on test score. a post-hoc test revealed that the \textit{Personalized} condition (M=7.30, SD=1.45) had significantly higher scores than both the \textit{Generalized} condition (M=5.85, SD=1.35), $p<0.01$, and the \textit{Textbook} condition (M=5.95, SD=1.57), $p=0.01$.  
    \item \textbf{Psychology}: We found a significant main effect of \textit{content style} on test score, $F(2,57)=3.17, p =0.04$. A post-hoc test revealed that there were no significant pairwise comparisons, however the \textit{Personalized} condition (M=5.80, SD=1.24) trended as having higher scores than both the \textit{Generalized} condition (M=4.75, SD=1.62), $p=0.07$, and the \textit{Textbook} condition (M=4.8, 1.57), $p=0.09$. 
    \item \textbf{Government}: we did not find a significant main effect of \textit{content style} on test score, $F(2,57)= 2.96, p =0.06$. Thus, no post-hoc tests were conducted.  
\end{itemize}
H2 was partially supported, as the \textit{Personalized} condition led to increased learning outcomes compared to the \textit{Generalized} and \textit{Textbook} conditions in two of the subjects. However, we did not find that the \textit{Generalized} condition resulted in superior learning outcomes over the \textit{Textbook} condition in any case, unlike our hypothesis.

\subsection{Learning Experience}
Figure \ref{fig:attractiveness} shows plots of the scale factors for experience. 
\subsubsection{Attractiveness}
We did not find significant interaction effects between \textit{content style} and \textit{subject}. We found a significant main effect of \textit{content style}, $F(2, 171)=11.06, p<0.01$. A post-hoc test revealed that both the \textit{Personalized} condition (M=5.25, SD=1.06) and the \textit{Generalized} condition (M=5.25, SD=0.88) had significantly higher ratings than the \textit{Textbook} condition (M=4.52, SD=1.03), supporting H1. There was no main effect of \textit{textbook subject}.

\subsubsection{Stimulation}
We did not find any significant main effects or interaction effects of \textit{content style} or \textit{subject}. 

\begin{figure} \centering
  \includegraphics[width=5in]{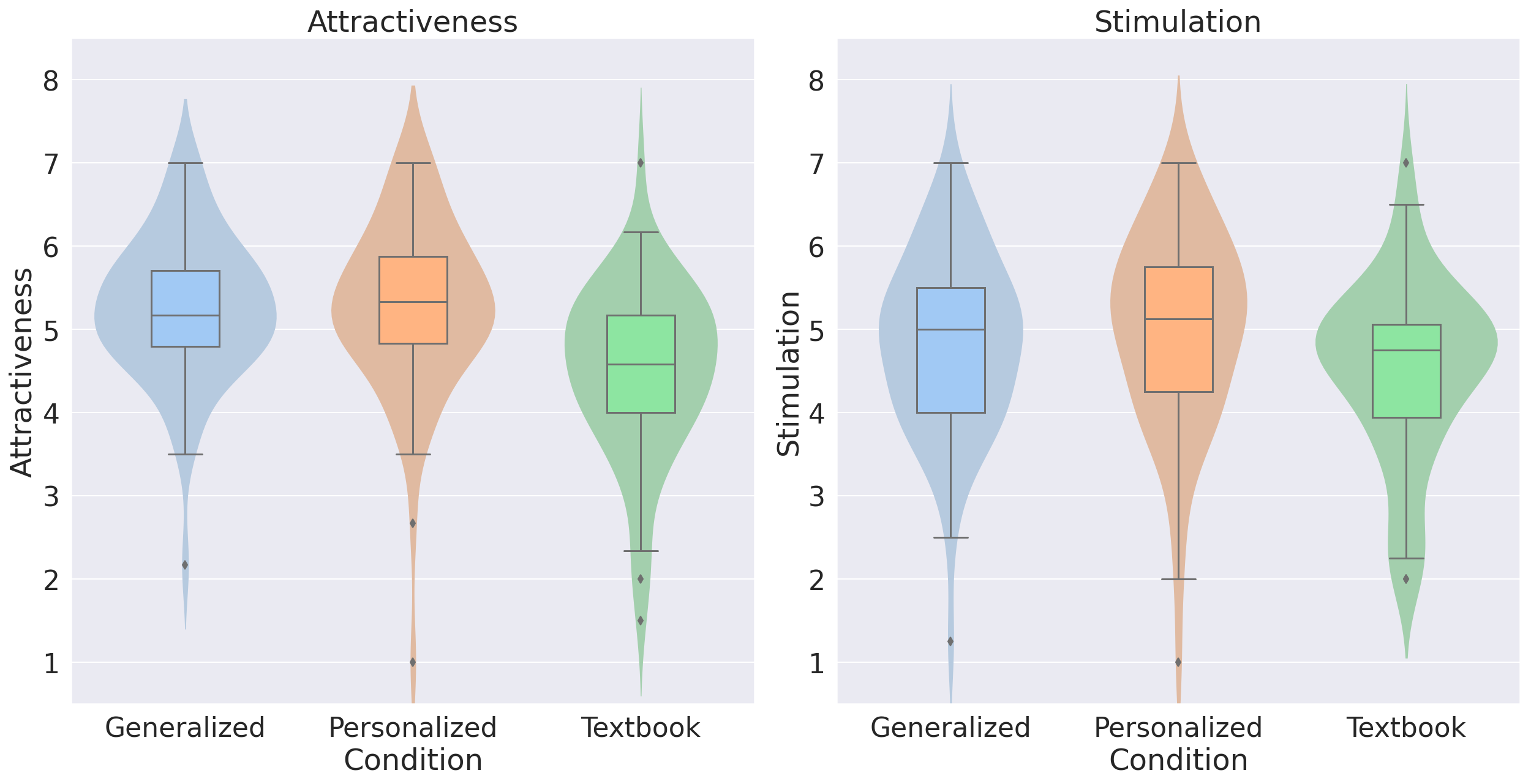}
  \caption{Violin and boxplots of Attractiveness and Stimulation scores of all content type conditions. Since there were no interaction effects of subject, all subjects are combined.}
  \Description{Violin and boxplots of Attractiveness and Stimulation ratings by condition. Since there were no interaction effects of subject, all subjects are combined. Both plots show that these ratings are generally higher for the podcast conditions.}
  \label{fig:attractiveness}
\end{figure}

\section{Discussion}
\subsection{AI-Generated Podcasts are Enjoyable (RQ1)}
Our findings showed that the podcast conditions received significantly higher \textit{Attractiveness} ratings compared to the textbook, irrespective of personalization. This is consistent with previous studies indicating that students generally prefer human-generated podcasts over textbooks \cite{evans2008, back2017}. Our results suggest that AI-generated podcasts can similarly replicate this engaging experience in real time and provide suitable supplements to textbooks.

Participants' qualitative responses highlighted a strong preference for the podcasts' entertainment value and casual nature. Notably, despite being unaware of the textbook condition, several participants commented on the perceived boredom associated with traditional learning materials and found value in the podcast format.

\begin{quote}
    \textit{"I liked the podcast setup of this. It was much more entertaining to learn about philosophy in this type of format rather than reading about it directly from a textbook."}
    \\ \\
    \textit{"One thing I liked is that it was actually enjoyable to listen to them, [unlike other materials]. I imagine I could listen to this while doing school work and actually learn something."}
\end{quote}

Students also appreciated the podcasts' casual and conversational style.

\begin{quote}
    \textit{"The podcast style of information made me feel like I was part of a conversation rather than being lectured to, which helped me understand the ideas discussed in a way that I wasn't expecting to."}
\\ \\ 
    \textit{"The language used was super informal, which honestly I thought was a plus because it made it feel like a quick informational session rather than [studying]. I think this would put more people at ease and make them more receptive."}
\end{quote}

Some students found the novel style of the podcast content initially uncomfortable, which they felt detracted from their learning experience. One student remarked, \textit{"I didn't like the style at first as it's very different from traditional lectures or textbooks, but it was easy to adapt to as it progressed."} This initial discomfort may have influenced the lack of significant differences in \textit{Stimulation} ratings. Nonetheless, \textit{Stimulation} ratings exhibited similar trends to \textit{Attractiveness} (see Figure \ref{fig:attractiveness}). Future improvements in the quality of AI-generated podcasts could enhance overall ratings and potentially elevate \textit{Stimulation} ratings to statistical significance.

\subsection{AI-Driven Personalization Can Improve Learning (RQ2/RQ3)} 
We found a significant main effect of \textit{content type} for the Philosophy and Psychology subjects, with the \textit{Personalized} condition showing higher test scores than both the \textit{Generalized} and \textit{Textbook} conditions (see Figure \ref{fig:test_scores}). However, no significant main effect was observed for the Government subject.

We hypothesize that this difference may stem from the subject's relevance to the user's major and interests. For example, many Philosophy and Psychology concepts can be applied to other fields of study or linked to pop culture references. Our qualitative results indicated that participants found these connections engaging, which ultimately enhanced their understanding of the material. Moreover, participants also found it helpful to see examples of how the subject related to their field of study. 

\begin{quote}
    \textit{"I liked the references to different things in order to explain the concept of philosophy. As someone who loves [storytelling games] and is also a coder, I felt that my understanding of philosophy was strengthened from the way they used those two concepts of making a game like [a particular story game] and coding to explain philosophy."}
\\ \\ 
    \textit{"I love learning new things, especially ones I have no prior knowledge about. I particularly enjoyed that the examples they made were more relatable, thus making the content itself [easy to learn]."}
\end{quote}

Bernacki and Walkington \cite{bernacki2018role} theorize that personalization can activate students' prior knowledge, potentially improving learning. Our findings suggest that AI-driven personalization can achieve effects consistent with this theory. However, in the case of the Government subject, several participants perceived the personalization as irrelevant, with some commenting that the connections made felt inappropriate and even distracting. This lack of relevance may explain why we did not observe increased test scores for this subject, indicating that AI-driven personalization might not be effective when the context lacks a clear connection to the user's interests and experiences.

\subsection{Personalization Has Minimal Impact on Enjoyment (RQ1)}
Interestingly, we found that personalization did not impact \textit{Attractiveness} ratings, contrary to our hypothesis. This suggests that the enjoyment of the experience was influenced more by the lesson format than by the content itself. While personalization enhanced the content with relatable terms or analogies, it may have primarily facilitated learning rather than contributing to the overall enjoyment of the experience. Additionally, \textit{Attractiveness} ratings did not correlate with test scores. Although students may have enjoyed the podcasts more than the textbook, this enjoyment did not necessarily translate into deeper learning. Alternatively, as Goldman suggests, podcasts may be more distracting \cite{goldman2018impact}, potentially offsetting any learning benefits gained from increased enjoyment.

Additionally, while users could choose how much information about themselves to provide, the system applied all available information without adjusting the level of personalization. Qualitative feedback indicated that some students desired more personalization (e.g., more references to their interests), while others preferred less. This lack of control over the granularity of personalization may have influenced \textit{Attractiveness} ratings. For instance, previous research has shown that interest-based personalization can sometimes reduce enjoyment \cite{iterbeke2022role, VANDEWEIJERBERGSMA2021}, which may explain the lack of significant difference between \textit{Personalized} and \textit{Generalized} podcasts.
 
\subsection{Recommendations for Designing AI-Generated Educational Content}
Participant feedback emphasized the need for visual elements in the content. Over half of those who provided specific suggestions requested visuals such as animations or slideshows to complement the podcasts. This points to a promising direction for future research, particularly with the rapid advancements in generative AI. As one participant noted, \textit{"I would love if there were little animations or pictures to help stay focused and make the lesson more exciting."} 

Participants noted that the TTS voices occasionally sounded unnatural and exhibited minor audio glitches. Given the importance of audio quality in podcasts, improving the audio models should be a priority. Future systems should also explore the effectiveness of multimodal generative lessons that integrate both audio and visual elements. For example, it is essential to determine whether slideshows, animations, or bullet points can enhance the effectiveness of generative podcasts. However, this must be balanced with the risk of overwhelming learners with excessive multimodal content, which could distract and impede learning \cite{mayer2005cognitive}. Conversely, animations or images might enhance engagement and support learning, according to our user feedback.

\subsection{Ethical Implications}
It is also important to consider the ethical concerns raised by some students. While the majority of participants responded positively to the AI-generated podcasts, some participants (5\% of total) explicitly stated their dislike for AI-generated content. Although representing a fairly small portion of the data, their presence highlights the importance of acknowledging potential user concerns and considering them in future design decisions.
 Among the 120 students assigned to the podcast conditions, approximately 5\% expressed a dislike of AI. For instance, one participant felt uneasy about the AI pretending to know them, despite being an artificial entity: 
\begin{quote}
    \textit{“It makes me believe it is written and voiced by AI, which is like when an email or essay is clearly made with ChatGPT, when someone pretends to put in effort where there is none, but passes it off as effort. Why would someone want to spend their time and effort reading or engaging with something or someone who chooses an audience he knows nothing about?"}
\end{quote}

Additionally, it is important to note that potential bias may also be a point of concern for AI-generated lessons. For example, Blanchard and Mohammed \cite{BlanchardMohammed2024} note that LLM technologies are primarily developed in the West and may exhibit inconsistencies in cultural personalities, creating challenges for globalizing such technology for education and potentially resulting in inequitable outcomes. While our work focused on students in the United States, the findings may not be applicable to students from other cultural contexts. It is crucial to explore how to develop and train models that accommodate diverse cultural perspectives.

\subsection{Limitations and Future work}
We acknowledge that this study's scope was limited, involving only a single textbook chapter per participant within a relatively short timeframe. Since students typically engage with multiple chapters over extended periods, future research should explore longer study durations to assess how learning and enjoyment evolve over time. Additionally, our study focused on three college-level textbooks and included only U.S. college students. Further investigation is needed to examine these effects across a broader range of subjects, demographics, and educational levels. Moreover, future work should explore interactive elements in AI-generated podcasts, as seen in prior research on AI-generated content \cite{laban2022} and educational technologies \cite{kharrufa2024}, to enhance personalization and learner engagement.

\section{Conclusion}
Our results demonstrate the potential of using large language models (LLMs) to transform traditional text-based content into more engaging formats, such as podcasts. Our findings indicate that U.S. college students generally preferred AI-generated podcasts over textbooks, appreciating the entertainment value of the format. Surprisingly, while personalization did not significantly enhance the overall enjoyment of the learning experience, it improved comprehension and learning outcomes in subjects where the personalized content aligned well with students' interests and prior experiences.

These findings suggest that AI-driven personalization can make educational content more relevant and effective, especially when tailored to the learner's context. Educators and content creators should consider integrating AI-generated podcasts into their teaching strategies to enhance engagement and comprehension. However, the varied impact of personalization across different subjects highlights the need for further investigation into how and when personalization should be applied.

Future research should explore the long-term effects of AI-generated podcasts on learning retention, the influence of various personalization strategies on different demographic groups, and the broader implications of AI-driven personalization in diverse educational settings. Additionally, it will be important to examine how AI-generated content can be integrated with other teaching modalities to create a more holistic and adaptable educational experience.

\bibliographystyle{ACM-Reference-Format}
\bibliography{sample-base}
\newpage
\appendix
\section{Participant Demographics}
\begin{table}[h!] \centering \small
\caption{Demographics of all participants, separated by condition and gender (woman, man, and non-binary).}
\begin{tabular}{|llll|}
\hline
\multicolumn{1}{|l|}{} & \multicolumn{1}{l|}{\textbf{Age}} & \multicolumn{1}{l|}{\textbf{Gender}} & \textbf{Major} \\ \hline
\multicolumn{4}{|c|}{{\ul \textit{Philosophy}}} \\ \hline
\multicolumn{1}{|l|}{\textbf{Generalized}} & \multicolumn{1}{l|}{26.5 (7.4)} & \multicolumn{1}{l|}{11W, 9M, 0NB} & \begin{tabular}[c]{@{}l@{}}Accounting (4), Biology (2), Journalism (1), Business (2), \\ Clinical Nutrition (1), Computer Science (4), Linguistics (1), \\ Marketing (1), Psychology (3), Sociology (1)\end{tabular} \\ \hline
\multicolumn{1}{|l|}{\textbf{Personalized}} & \multicolumn{1}{l|}{25.4 (6.4)} & \multicolumn{1}{l|}{11W, 9M, 0NB} & \begin{tabular}[c]{@{}l@{}}Accounting (1), Arts (1), Business (1), Computer Science (5), \\ Geology (1), Graphic Design (1), History (2), \\ Information Technology (2), Nursing (1),  Physics (1), \\ Psychology (2), Sociology (1), Sports Management (1)\end{tabular} \\ \hline
\multicolumn{1}{|l|}{\textbf{Textbook}} & \multicolumn{1}{l|}{25.9 (7.1)} & \multicolumn{1}{l|}{6W, 12M, 2NB} & \begin{tabular}[c]{@{}l@{}}Accounting (1), Aerospace Engineering (1), Business (1), \\ Computer Science (4), Finance (1), Graphic Design (1), \\ Mechanial Engineering (1), Middle Childhood Education (1), \\ Network and System Administration (1), Nursing (1),\\ Nutrition (1), Pre-Medicine (2), Psychology (4)\end{tabular} \\ \hline
\multicolumn{4}{|c|}{{\ul \textit{Psychology}}} \\ \hline
\multicolumn{1}{|l|}{\textbf{Generalized}} & \multicolumn{1}{l|}{29.0 (8.1)} & \multicolumn{1}{l|}{11W, 9M, 0MB} & \begin{tabular}[c]{@{}l@{}}Accounting (2), Business (2), Business Administration (1)\\ Business Management (4), Computer Science (3)\\ Dental Hygiene (1), Engineering (1), Gender Studies (1)\\ Graphic Design (1), History (1), Hospitality (1), Humanities (1)\\ Nursing (1)\end{tabular} \\ \hline
\multicolumn{1}{|l|}{\textbf{Personalized}} & \multicolumn{1}{l|}{29.8 (7.6)} & \multicolumn{1}{l|}{10W, 9M, 1NB} & \begin{tabular}[c]{@{}l@{}}Administration of Justice (1), Biology (1), Business (1)\\ Communications (1), Computer Science (3), Education (2)\\ Geoscience (1), History (3), International Business (1)\\ Liberal Arts (1), Marketing (1), Math (1), Multimedia (1)\\ Music Performance (1), Studio Art (1)\end{tabular} \\ \hline
\multicolumn{1}{|l|}{\textbf{Textbook}} & \multicolumn{1}{l|}{26.0 (6.8)} & \multicolumn{1}{l|}{9W, 11M, 1NB} & \begin{tabular}[c]{@{}l@{}}Biology (1), Business (2), Business Administration (2)\\ Computer Science (4), Economics (1), Environmental Science (1)\\ Electrical Engineering (1), Finance (1), General Studies (1)\\ Graphic Design (1), History (1), Information Technology (2)\\ Law \& Society (1), Liberal Arts (1)\end{tabular} \\ \hline
\multicolumn{4}{|c|}{{\ul \textit{U.S Government}}} \\ \hline
\multicolumn{1}{|l|}{\textbf{Generalized}} & \multicolumn{1}{l|}{26.5 (9.0)} & \multicolumn{1}{l|}{12W, 8M, 0NB} & \begin{tabular}[c]{@{}l@{}}Biotechnology (1), Business (3), Business Administration (1)\\ Communications, Computer Engineering (1), Computer Science (2)\\ Economics (1), Exercise Science (1), General Studies (1), Health (2)\\ Family Studies (1), Hospitality (1), Psychology (3)\end{tabular} \\ \hline
\multicolumn{1}{|l|}{\textbf{Personalized}} & \multicolumn{1}{l|}{27.7 (8.1)} & \multicolumn{1}{l|}{10W, 10M, 0NB} & \begin{tabular}[c]{@{}l@{}}Aerospace Engineering (1), Bioengineering (1), Biology (1)\\ Business (1), Accounting (1), Chemistry (1), Communication (1)\\ Computer Science (2), Finance (1), Humanities (1), \\ Information Technology (1), Law (1), Marketing (1),\\ Marriage and Family Therapy (1), Math (1),\\ Mechanical Engineering (1), Neuroscience (1), Psychology (2)\end{tabular} \\ \hline
\multicolumn{1}{|l|}{\textbf{Textbook}} & \multicolumn{1}{l|}{25.9 (6.4)} & \multicolumn{1}{l|}{11W, 9M, 0NB} & \begin{tabular}[c]{@{}l@{}}Arts (1), Biochemistry (1), Biology (4), Business Administration (1)\\ Computer Engineering (1), Computer Science (1), English (1),\\ Finance (1), Graphic Design (2), Health Science (2) History (1), \\ Information Technology (1), Nursing (1), Religion (1), \\ Theatre Tech and Design (1)\end{tabular} \\ \hline
\end{tabular}
\label{table:participants}
\end{table}
\newpage
\section{Results of All Statistical Tests}
\begin{table}[h!] \centering
\caption{Summary of two-way ANOVA tests for main and
interaction effects of content style and subject. Bolded values denotes a significant difference at p < 0.05. Italics signify a significant main effect, but the factor has an involvement in an interaction effect.}
\begin{tabular}{@{}|l|cc|cc|cc|@{}}
\hline
 & \multicolumn{2}{c|}{\textbf{Content style}} & \multicolumn{2}{c|}{\textbf{Subject}} & \multicolumn{2}{c|}{\textbf{\begin{tabular}[c]{@{}c@{}}Content style x\\ Subject\end{tabular}}} \\ \hline
 & \textbf{F} & \textit{\textbf{p}} & \textbf{F} & \textit{\textbf{p}} & \textbf{F} & \textit{\textbf{p}} \\ \hline
\textbf{Test score} & \textit{7.86} & \textit{\textless{}0.01} & \textit{19.27} & \textit{\textless{}0.01} & 2.27 & \textbf{\textless{}0.01} \\
\textbf{Attractiveness} & 11.06 & \textbf{\textless{}0.01} & 0.21 & 0.81 & 1.41 & 0.23 \\
\textbf{Stimulation} & 2.49 & 0.08 & 0.84 & 0.43 & 0.93 & 0.45 \\ \hline
\end{tabular}
\label{table:anova}
\end{table}
\end{document}